\documentclass[epj]{svjour}
\usepackage{graphics}
\usepackage{times}
\usepackage{hyperref}
\usepackage{amsmath}
\usepackage{algpseudocode}

\begin{document}

\title{A strategy to suppress recurrence in grid-based Vlasov solvers}
\subtitle{}
\author{Lukas Einkemmer\inst{1}\hskip-1pt$^,$%
\thanks{This work is supported by the Fonds zur F\"orderung der Wissenschaften (FWF) -- project id: P25346.\newline e-mail: lukas.einkemmer@uibk.ac.at}%
\and Alexander Ostermann\inst{1}%
}                     
\institute{Department of Mathematics, University of Innsbruck, Technikerstra\ss e 19a, A-6020 Innsbruck, Austria}
\date{Received: date / Revised version: date}
%
\abstract{
In this paper we propose a strategy to suppress the recurrence effect present in grid-based Vlasov solvers. This method is formulated by introducing a cutoff frequency in Fourier space. Since this cutoff only has to be performed after a number of time steps, the scheme can be implemented efficiently and can relatively easily be incorporated into existing Vlasov solvers. Furthermore, the scheme proposed retains the advantage of grid-based methods in that high accuracy can be achieved. This is due to the fact that in contrast to the scheme proposed by Abbasi et al. no statistical noise is introduced into the simulation. We will illustrate the utility of the method proposed by performing a number of numerical simulations, including the plasma echo phenomenon, using a discontinuous Galerkin approximation in space and a Strang splitting based time integration.
} 
\maketitle
%

\section{Introduction}

A plasma can be modeled using a particle-density $f(t,\boldsymbol{x},\boldsymbol{v})$ that obeys the Vlasov equation (this is the so-called kinetic description of a plasma). However, plasma phenomena inherently have to take electromagnetic effects into account. Therefore, the Vlasov equation is coupled to an appropriate model of the electromagnetic field. In the non-relativistic and electrostatic regime the so-called Vlasov--Poisson equations (a system of two coupled partial differential equations) \begin{align} \label{eq:vp}
	\partial_t f(t,\boldsymbol{x},\boldsymbol{v}) + \boldsymbol{v} \cdot \nabla_{\boldsymbol{x}} f(t,\boldsymbol{x},\boldsymbol{v}) +
	\boldsymbol{E} \cdot \nabla_{\boldsymbol{v}} f(t,\boldsymbol{x},\boldsymbol{v})=0, \\
	\nabla_x \cdot \boldsymbol{E} = \rho,
\end{align}
are usually the starting point for numerical simulations. We denote the position by $\boldsymbol{x}$ and the velocity by $\boldsymbol{v}$. The dimensionless Vlasov--Poisson equations \eqref{eq:vp} have to be supplemented by the quasi-neutrality condition; furthermore, appropriate boundary and initial conditions have to be specified. Note that the particle-density $f$ is stated in an up to $3+3$ dimensional phase space.

Due to the high dimensional setting of the equations considered, the most common numerical approach are so-called particle methods. In this class of methods, the phase space is left to be continuous and a (large) number of particles with various starting points are advanced in time. This is possible due to the structure of the equations, which implies that a single particle evolves along a trajectory given by an ordinary differential equation. A number of such methods have been developed; most notably, the particle-in-cell (PIC) method. Particle methods have been extensively used for various applications (see e.g.~\cite{fahey2008}). The PIC scheme gives reasonable results in case where the tail of the distribution is negligible. If this is not the case the method suffers from numerical noise that only decreases as $1/\sqrt{n}$, where $n$ denotes the number of particles (see e.g.~\cite{filbet2003}). Motivated by these considerations, a number of schemes employing discretization in phase space have been proposed.

In these grid-based methods the computational domain is subdivided into a number of cells on which the particle-density is approximated by finite differences, finite elements, finite volumes, or a discontinuous Galerkin method (see e.g.~\cite{mangeney2002} and \cite{heath2012}). The time integration is usually accomplished by recognizing that a splitting of the Vlasov equation results in two parts the solution of which can be written down analytically as a translation in phase space (i.e., the characteristics can be computed analytically). Following the seminal work of Cheng \& Knorr \cite{cheng1976} this exact solutions are then combined by Strang splitting to yield a scheme of order two. This procedure yields an efficient scheme for the solution of the Vlasov--Poisson equations.

Unfortunately, however, an approximation that employs a grid of constant size in the velocity direction does exhibit the so-called recurrence phenomenon. This purely numerical artifact is understood by considering that for a piecewise constant approximation in velocity space the numerical solution returns to the initial value after a time $t_{\mathrm{r}}\propto 1/h$, where $h$ denotes the cell size. Before proceeding let us make the following remarks. First, the recurrence time $t_{\mathrm{r}}$ can be pushed arbitrarily far into the future; however, doing so requires a large number of grid points which is prohibitive from an efficiency standpoint. Second, this behavior is, somewhat surprisingly, also a problem in higher order space discretization. Third, PIC methods do not suffer from this phenomenon due to the randomization of the phase space. However, since the high accuracy of the grid based solvers is required in many application, designing a recurrence free method is of interest.

This paper proceeds as follows. In section \ref{sec:recurrence} the recurrence phenomenon is explained in more detail. Furthermore, it is discussed why the randomization method recently proposed in \cite{abbasi2011} is inherently limited. In section \ref{sec:method} we then outline the method proposed in this paper which is based on the control of high spatial frequencies.\footnote{In the following we are exclusively concerned with (spatial) frequencies, i.e. wavenumbers. Thus, we will omit the prefix spatial and use the letter $k$ to denote spatial frequencies/wavenumbers in mathematical expressions.} Finally, we conduct a number of numerical simulations in case of the Vlasov--Poisson equations in section \ref{sec:numerics}.

\section{Recurrence}\label{sec:recurrence}

The recurrence phenomenon can be explained most easily for an advection equation that is formulated in a $1+1$ dimensional phase space; that is,
\begin{equation} \label{eq:advection}
	\partial_t f(t,x,v) = v \partial_x f(t,x,v).
\end{equation}
The exact solution of equation \eqref{eq:advection} can simply be written down as follows
\[ f(t,x,v) = f(0,x-vt,v). \]
Now, for the sake of concreteness, let us assume that the initial value is given by
\begin{equation} \label{eq:initial}
	f(0,x,v) =  \cos(k x) \mathrm{e}^{-\beta v^2},
\end{equation}
i.e., we assume an equilibrium density in velocity space that is modulated by an oscillation in position space. The exact solution of \eqref{eq:advection} is then given by
\begin{equation} \label{eq:solution}
	f(t,x,v) =  \cos(k (x-vt)) \mathrm{e}^{-\beta v^2}.
\end{equation}
Now, a quantity of interest especially in the context of Landau damping is the energy stored in the electric field. In the simplified model no electric fields are considered, even so we can compute the electric energy $\mathcal{E}$ (which we consider as an auxiliary quantity) and find that
\begin{equation} \label{eq:electric_energy_advection}
	\mathcal{E} \propto \mathrm{e}^{-\frac{k^2}{2\beta} t^2}.
\end{equation}
That is, the electric energy decays exponentially in this configuration.

Let us assume that we are able to devise a numerical scheme to compute the exact solution of \eqref{eq:advection} but that we are only able to store the values at an equidistant grid $v_i = i h$; that is, only the values $f(t,x,v_i)$ are available. Due to the structure of the grid, we immediately follow from \eqref{eq:solution} that after a time $t=2\pi/(k h)$ the values $f(t,x,v_i)$ are indistinguishable from $f(0,x,v_i)$ and thus all derived quantities (such as the electric energy) are equal as well. However, this observation is incompatible with \eqref{eq:electric_energy_advection} which predicts an exponential decay of the electric energy. This phenomenon is called recurrence. The recurrence time $t_r=2\pi/(kh)$ is inversely proportional to the cell size $h$. The same recurrence can be observed for the Vlasov--Poisson equations (see \cite{einkemmer2012}).

\begin{figure}[h]
	\centering
	\includegraphics{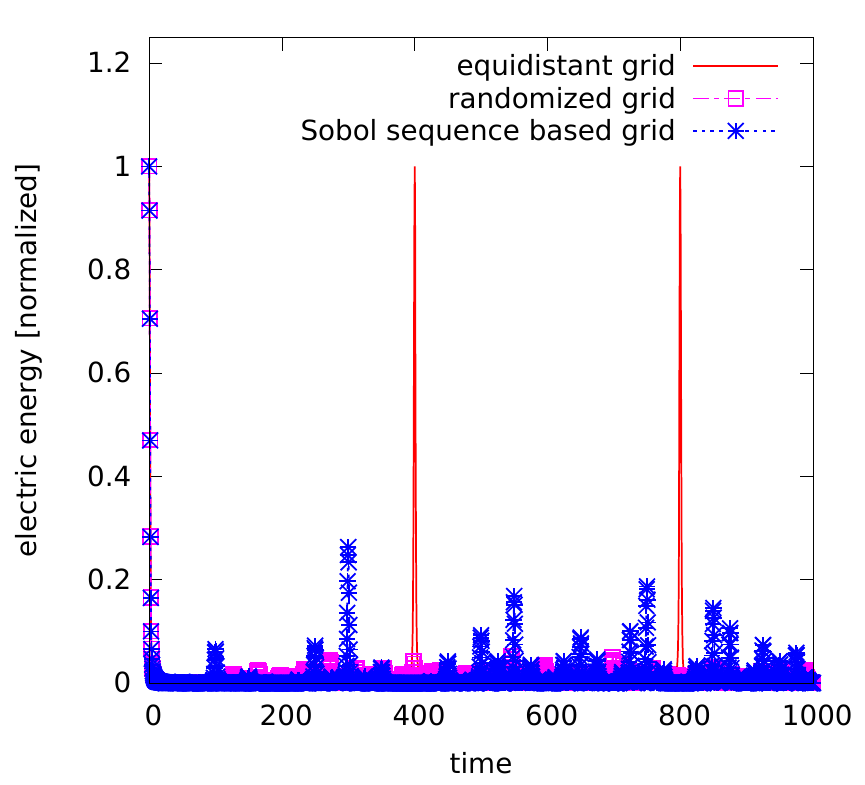}
	\caption{The normalized (to the initial value) electric energy is shown
as a function of time for $400$ grid points in both space and velocity on the domain $[-1,1]\times[0,1]$ and for $k=2\pi$, $\beta=20$. The single phase space point inside each cell is randomized or determined by a Sobol sequence, respectively.}
	\label{fig:simpleadvection}
\end{figure}

The following remedy has been proposed in \cite{abbasi2011}:  one randomizes the phase-point velocities in each cell and then computes the desired values by interpolation. This strategy results in a clear reduction of the recurrence phenomenon (see Figure \ref{fig:simpleadvection}). For comparison, we have also implemented a pseudo-randomization based on Sobol sequences (generated by the software described in \cite{joe2003}) and this seems to give better results for some time points; however, it also introduces significantly larger spikes which limits its practical applicability  (see Figure~\ref{fig:simpleadvection}).

In \cite{abbasi2011} multiple points are randomized within a single cell and a reduction in the amplitude of the recurrence of about a factor of $20$ is observed for the linear Landau damping. Note, however, that in this scenario the reduction in the amplitude scales only as $1/\sqrt{N}$ due to the randomization, where $N$ is the number of discretization points in the velocity direction. Thus, if high accuracy is desired (such as in the plasma echo application discussed in section \ref{sec:numerics}) a grid size has to be chosen which is computationally prohibitive. Furthermore, using this approach requires significant modifications in order to integrate it into an already existing Vlasov solver. In the next section we will propose a new strategy that remedies both of these shortcomings.

\section{Description of the proposed method} \label{sec:method}

In this paper we take the viewpoint that the recurrence effect is not a consequence of the lack of accuracy of the underlying space discretization, but a result of aliasing. Due to the creation of higher and higher frequencies in phase space aliasing of the high frequencies introduces an error in the macroscopic quantities (such as the electric energy) that are usually computed by integrating over some subset of the phase space. This behavior can be observed from the time evolution present in \eqref{eq:solution}.

Since these high frequencies develop gradually while integrating the partial differential equations forward in time, we propose to introduce a cutoff that removes (sets to zero) high frequencies in Fourier space. For this procedure we introduce two parameters: the cutoff frequency $k_{\mathrm{cutoff}}$ and the cutoff time $t_{\mathrm{cutoff}}$. A single time step within the algorithm then proceeds as illustrated by the following pseudocode
\begin{algorithmic}
\If {$t_n-t_{\mathrm{last}} > t_{\mathrm{cutoff}}$}
	\State $t_{\mathrm{last}}=t_n$
	\For {$i$}
		\State $g=f(t_n,x_i,\cdot)$
		\State $g_F={}$\Call{Fourier}{$g$}
		\For {$k=k_{\mathrm{cutoff}} : k_{\mathrm{max}}$}
			\State $g_F[k]=0$
		\EndFor
		\State $f(t_{n},x_i,\cdot)={}$\Call{Inverse\_Fourier}{$g_F$}
	\EndFor
\EndIf
\State $f(t_{n+1},\cdot,\cdot)={}$\Call{Regular\_Timestep}{$f(t_n,\cdot,\cdot)$,$\tau$}
\State $t_{n+1} = t_n + \tau$
\end{algorithmic}
where the implementation details depend on the specific fast Fourier transform (FFT) used in the implementation.

Note, however, that the above implementation can only be directly applied if a space approximation is used that stores a single value in each cell. In recent years discontinuous Galerkin (dG) approximations have become popular as a tool to solve hyperbolic partial differential equations (see, for example, \cite{einkemmer2012}, \cite{heath2012}, and \cite{rossmanith2011}). In such methods a number of coefficients are stored for each cell representing the coefficients of the expansion in some set of orthonormal basis functions (such as Legendre polynomials). The same remark applies to higher-order finite element methods.

In the present discussion we are most interested in the discontinuous Galerkin method described in \cite{einkemmer2012}. There are at least two generalizations to the procedure outlined in this section. First, we could, for a fixed position $x$, compute a number of equidistant values of $f(x,v_j)$ where the points $v_j$ do not coincide with the cell boundaries. However, due to the discontinuous Galerkin approximation, the particle-density $f$ is discontinuous along the cell edges. This, unfortunately, introduces artificial peaks in the spectrum at multiples of the maximal frequency that can be resolved by using a single data point per cell only; thus, the problem actually becomes worse, if the order of the dG method is increased. Furthermore, this method is relatively expensive as we have to compute a large number of function evaluations.

The alternative method we consider here is to treat the different coefficients as separate entities. Thus, given $K$ coefficients we compute $K$ FFTs of length $N$ using the vector of the value of the coefficients in each cell (where $N$ denotes the number of cells used in the velocity direction). This has the advantage that all the values are already stored in memory (and therefore no computation in addition to the FFT transforms has to be done). Furthermore, no artifacts are introduced in the spectrum. The numerical simulation conducted in the next section confirm that this methods in fact does work well.

Finally, let us discuss how to choose the cutoff frequency $k_{\mathrm{cutoff}}$ and the time span $t_{\mathrm{cutoff}}$ after which a cutoff should be performed. The requirement here is that no frequency can surpass the highest frequency represented in computer memory. Thus, let us assume that the highest frequency in the initial value is $k$ in the space and $k_1$ in the velocity direction. Then an advection modifies a pure wave $\cos (k x)\cos (k_1 v)$ at time $0$ to yield $\cos(k(x-v t)) \cos (k_1 v)$ at time $t$. Thus, the highest frequency in the velocity direction is equal to $kt+k_1$. Note that the fundamental frequency in Fourier space is $k_{0}=2\pi/(2V)$, where $V$ represents the highest possible velocity that is used in the approximation. Therefore, we argue that the increase in frequency $k t$ should be smaller than $m k_0$, i.e. the frequency range we cut off. Solving for the time gives
\begin{equation} \label{eq:cutoff}
	t_{\mathrm{cutoff}} < \frac{m \pi}{V k},
\end{equation}
where $m$ is the number of frequency components that are set to zero. Note that the cutoff frequency is given by $k_{\mathrm{cutoff}}=(M-m)k_0$, where $M$ is the number of Fourier modes stored in memory. In our numerical experiments we achieved the best results by using an $m$ that represents between $3$ and $15$ percent of the spectrum. Clearly, both $m$ and $t_\mathrm{cutoff}$ should not be chosen too small as it is advantageous to cut off a reasonable number of cells in order to exploit averaging of the error introduced by this procedure. On the other hand, the argument conducted here only takes into consideration the linear case, where a single frequency is present in the initial value. Due to non-linear effects higher frequencies can be generated in the space direction during the course of the time evolution. Thus, in practical simulations we have to choose a value for $t_{\mathrm{cutoff}}$ that is significantly smaller than $m\pi/(Vk)$ in order to achieve optimal performance.

\section{Numerical examples} \label{sec:numerics}

\begin{figure*}
	\centering
	\includegraphics{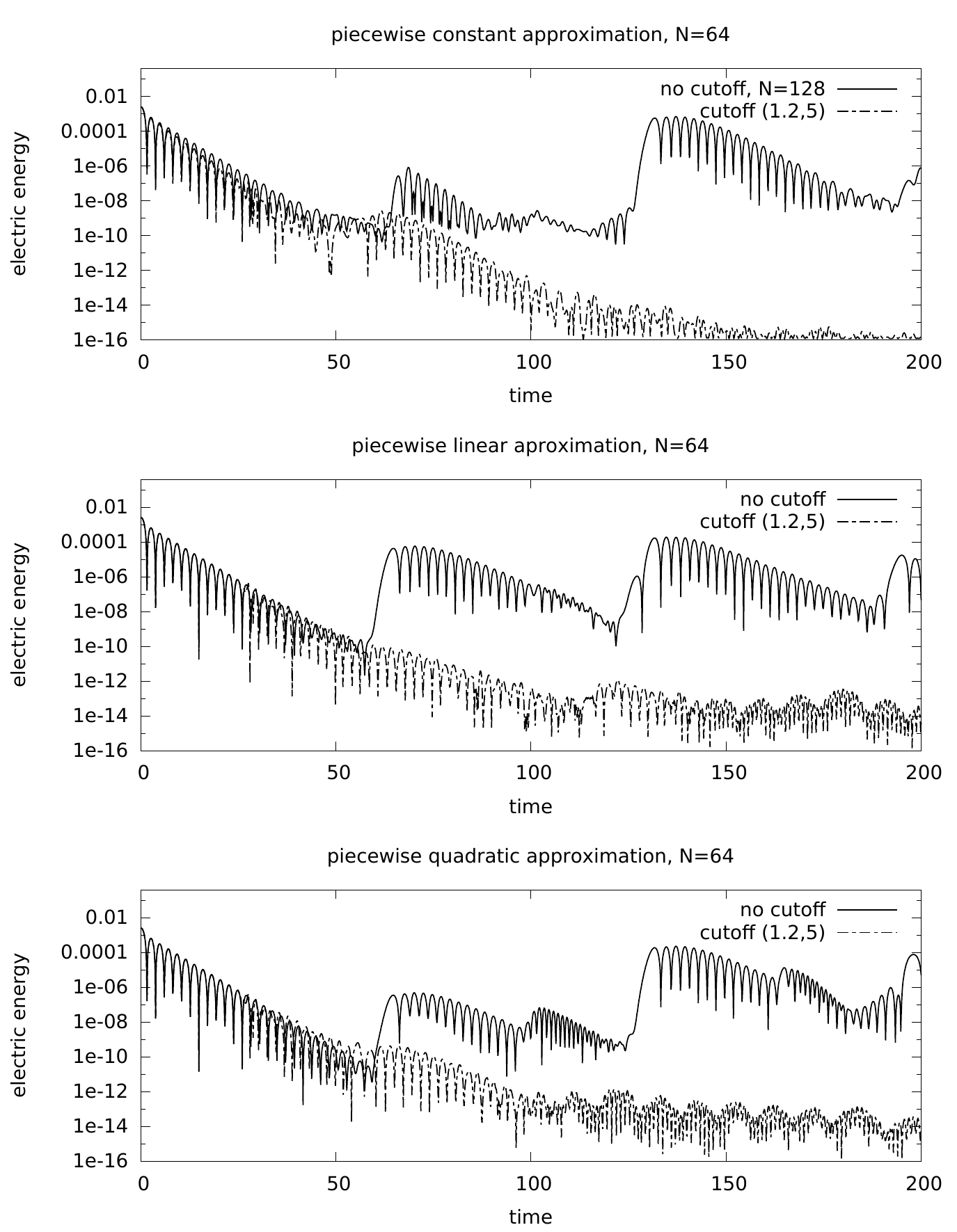}
	\caption{The numerical solution of the $1+1$ dimensional Vlasov--Poisson equations for the linear Landau damping test case (with $\alpha=0.01$) is shown. We employ a discontinuous Galerkin approximation with $64$ grid points (with the exception of the top figure where $128$ grid points are employed in the case where no cutoff is performed) in both the space and velocity direction and piecewise constant (top), piecewise linear (middle), and piecewise quadratic (bottom) polynomials. The time integration is performed using a Strang splitting scheme with a step size equal to $\tau=0.1$. The first parameter (displayed in parentheses in the plot) determines the time span after which a cutoff is performed and the second parameter gives the number of the frequency components that are set to zero. The maximum resolved velocity is equal in magnitude to $V=6$ and periodic boundary conditions are imposed.  This implies a recurrence time $t_r=2 \pi/(k h) \approx 67.02 $.}
	\label{fig:linear-landau}
\end{figure*}

\begin{figure*}
	\centering
	\includegraphics{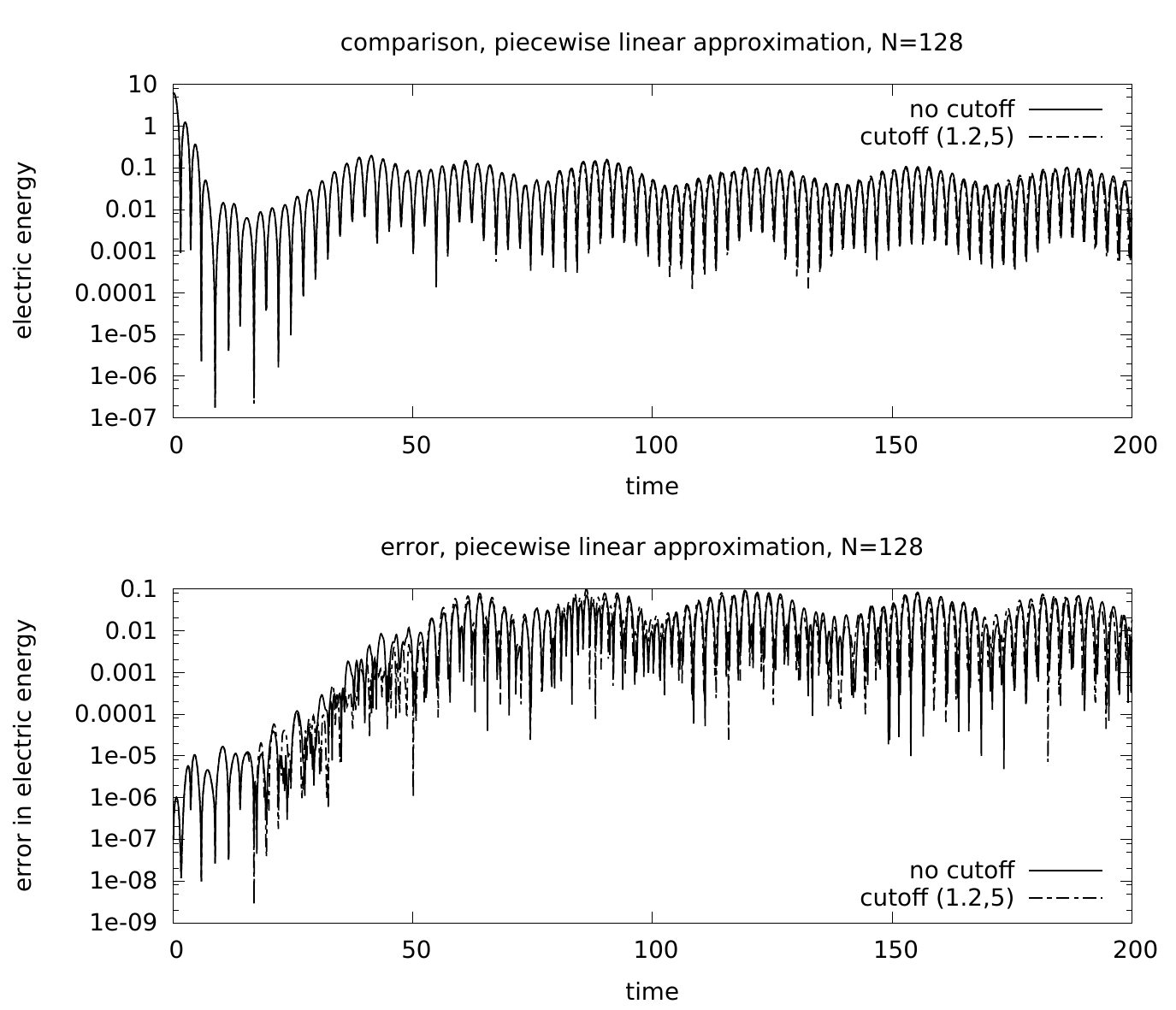}
	\caption{The numerical solution of the $1+1$ dimensional Vlasov--Poisson equations for the nonlinear Landau damping test case (with $\alpha=0.5$) is shown. We employ a discontinuous Galerkin approximation with $128$ grid points in both the space and velocity direction and piecewise linear polynomials. The time integration is performed using a Strang splitting scheme with a step size equal to $\tau=0.1$. The first parameter (displayed in parentheses in the plot) determines the time span after which a cutoff is performed and the second parameter gives the number of frequency components that are set to zero. The maximum resolved velocity is equal in magnitude to $V=6$ and periodic boundary conditions are imposed. The top figure shows the time evolution of the electric energy while the bottom figure displays the error with respect to a reference solution (for which a sufficiently small time step and a sufficiently fine space discretization has been chosen). }
	\label{fig:nonlinear-landau}
\end{figure*}

The most basic test example to investigate the effect of recurrence for the Vlasov--Poisson equation is the linear Landau damping; that is, we assume an initial value
\[ f_0(x,v) = \frac{1}{\sqrt{2\pi}}\mathrm{e}^{-v^2/2} (1+\alpha \cos 0.5 x), \]
where $\alpha \ll 1$. Thus, we use a Maxwellian equilibrium that is disturbed by a small perturbation in the $x$-direction. In this case we know that the electric energy does decay exponentially with a decay rate $\gamma$ given by $\gamma\approx 0.1533$ (see, for example, \cite{crouseilles2011}).

However, due to the recurrence phenomenon the numerical solution eventually returns to the initial value after which the same decay is observed again. If a higher order space discretization is used (see the bottom two plots in Figure \ref{fig:linear-landau}), the amplitude of the first recurrence is somewhat suppressed, but the numerical solution almost returns to the initial value for any further multiple of the recurrence time $t_{\mathrm{r}}$. Furthermore, we can see from Figure \ref{fig:linear-landau} that after a cutoff is introduced, as described in section \ref{sec:method}, the recurrence effect can no longer be observed. This is true for piecewise linear, piecewise constant, and piecewise quadratic space approximations as shown in the figure.
For the simulations conducted, we have chosen to cut off the highest $5$ of the $33$ (positive) frequencies present in the spectrum. In accordance with the discussion in the previous section, we choose $t_{\mathrm{cutoff}}= 1.2$.

Note that achieving this result for the linear Landau damping case is not particularly difficult to accomplish. For example, just estimating the recurrence time and extrapolating the decay would yield similar results as those shown in Figure \ref{fig:linear-landau}. Therefore, we now use the nonlinear Landau damping to verify that a physical recurrence of the electric energy is not damped away by our method. For this numerical experiment we use the same initial value as in the linear Landau damping case but choose $\alpha=0.5$ (i.e. a perturbation that is close to unity). The electric energy then decays but after a relatively short time a growth phase sets in. The remaining evolution is then dominated by an oscillation in the amplitude of the electric field (see, for example, \cite{manfredi1997}).

This behavior is in fact observed in the numerical results shown in Figure \ref{fig:nonlinear-landau}. We compare the method proposed in this paper to the method where no cutoff is performed: there is a discernible difference in the value at a given time $t$. However, these differences are clearly within the tolerance of the space and time discretization used (see bottom of Figure \ref{fig:nonlinear-landau}).

Finally, let us discuss the application of the method proposed in this paper to the plasma echo phenomenon. In this instance, we excite two perturbations in the plasma at different times, both of which are damped away by Landau damping; however, after some time the amplitude of the electric field does show a growth phase followed by a decay (this reappearance is called the plasma echo). Plasma echoes of different amplitude can be observed at different times depending on the frequencies and separation of the original excitations. This phenomenon has been considered in some detail in the literature (see, for example,  \cite{gould1967} for the initial analytical analysis and \cite{hou2011} for a numerical treatment). We consider the following initial value in this discussion:
\[ f_0(x,v) = \frac{1}{\sqrt{2 \pi}} \mathrm{e}^{-v^2/2} (1+ \alpha\cos (k_1 x) ), \]
which corresponds to an excitation with wavenumber $k_1$ at time $t=0$. Furthermore, at time $t=t_2$ we excite a second perturbation with wavenumber $k_2$; that is, we superimpose
\[ \frac{\alpha}{\sqrt{2 \pi}} \mathrm{e}^{-v^2/2} \cos (k_2 x) \]
on the (numerical) solution at time $t_2$.

\begin{figure*}
	\centering
	\includegraphics{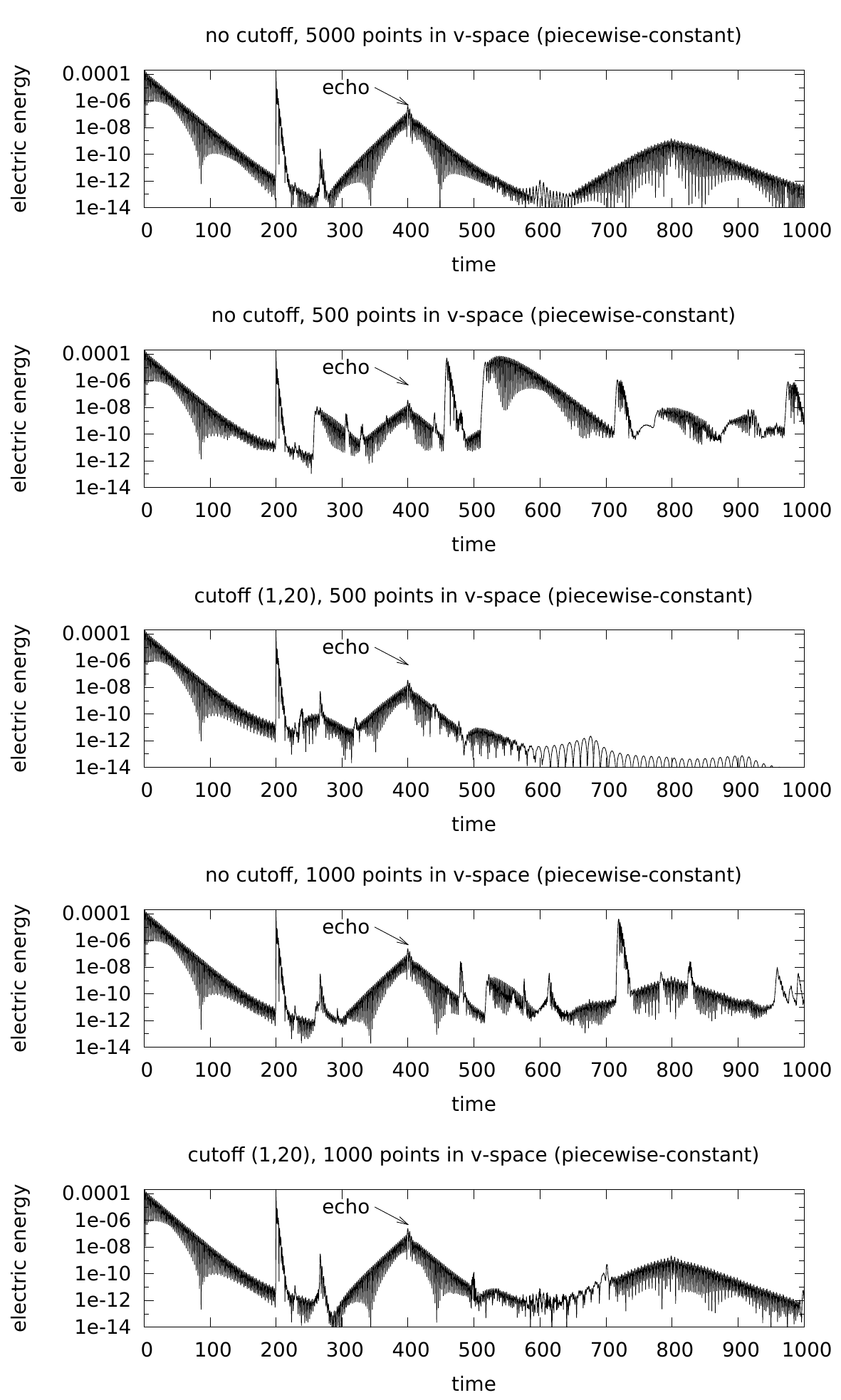}
	\caption{The numerical solution of the $1+1$ dimensional Vlasov--Poisson equations for the plasma echo problem (with $\alpha=10^{-3}$, $k_1=12 \pi/100$, $k_2=25 \pi/100$, and $t_2=200$) is shown. We employ a piecewise constant discontinuous Galerkin method in space; the time integration is performed using a Strang splitting scheme with a step size equal to $\tau=0.1$. The first parameter (displayed in parentheses in the plot) determines the time span after which a cutoff is performed and the second parameter gives the number of frequency components that are set to zero. The maximum  resolved velocity is equal in magnitude to $V=8$ and periodic boundary conditions are imposed on the domain $[0,100]$ in the $x$-direction. For this configuration the theory predicts a plasma echo at $t=400$ and $t=800$ which is in line with the numerical simulations shown here.}
	\label{fig:echo}
\end{figure*}

For such simulations the recurrence effect is a serious concern. Thus, space discretization in excess of $5000$ cells in the velocity direction have been employed to obtain reasonably accurate results (see \cite{hou2011}). This is necessary as the recurrence time must be well outside the computational domain. This is shown in Figure \ref{fig:echo}. At the top the reference solution (using $5000$ cells in both the space and velocity direction) is shown. In addition, we clearly observe that the solution with both $500$ and $1000$ cells (the second and fourth plot from the top, respectively) is completely spoiled by artifacts introduced by the recurrence phenomenon. In the $1000$ cells case the first echo is resolved reasonably well whereas in the $500$ cell case both echoes are completely buried in the recurrence. Furthermore, let us note that the amplitude of the plasma echo is quite small; that is, the randomization of phase space would not give reasonable results as it does not resolve peaks for which the amplitude is decreased by at least $3$ orders of magnitude compared to the initial value.

In the simulations we have conducted, $20$ frequencies are cut off (as we use a considerably larger number of cells than in the previous examples). The cutoff span $t_{\mathrm{cutoff}}$ is once again chosen as outlined in section \ref{sec:method} and yields a value close to unity. The simulation with $500$ cells (see third plot from the top in Figure \ref{fig:echo}) resolves the first plasma echo quite well (although the amplitude is diminished somewhat); however, the second plasma echo is not discernible. This is an accuracy problem but it demonstrates that no spurious recurrence effects are introduced. The simulation with $1000$ cells (see the bottom plot of Figure \ref{fig:echo}) is able to resolve both the first and second plasma echo (both in time as well as in amplitude). Therefore, we conclude that to resolve the plasma echo using the method proposed in this paper requires a significantly smaller number of cells in the velocity direction.

\section{Conclusion}

We have demonstrated that the algorithm proposed in this paper can result in a significant reduction in the number of cells that have to be employed in the context of numerical simulation using the Vlasov--Poisson equations, if recurrence is a concern. The plasma echo simulation demonstrates that this is of merit in simulations that are of interest by the plasma physics community. Furthermore, compared to the scheme developed in \cite{abbasi2011}, our method can be more easily incorporated into existing code bases and retains the high accuracy that is expected from grid based Vlasov solvers.

\bibliographystyle{plain}
\bibliography{recurrence}

\end{document}